# CMS Tracker Alignment Activities during LHC Long Shutdown 2

**Sandra Consuegra Rodríguez**[a,*]

[a] *Deutsches Elektronen-Synchrotron,*
*Notkestraße 85, 22607 Hamburg, Germany*

*E-mail:* sandra.consuegra.rodriguez@desy.de

The strategies for and the performance of the CMS tracker alignment during the 2021-2022 LHC commissioning preceding the Run 3 data-taking period are described. The results of the very first tracker alignment after the pixel reinstallation, performed with cosmic ray muons recorded with the solenoid magnet off are presented. Also, the performance of the first alignment of the commissioning period with collision data events, collected at center-of-mass energy of 900 GeV, is presented. Finally, the tracker alignment effort during the final countdown to LHC Run 3 is discussed.

[1] On behalf of the CMS Collaboration
*Speaker

## 1. CMS tracker detector

The innermost tracking system of the CMS experiment, called the tracker, consists of two devices, the Silicon Pixel and Silicon Strip detectors, within a length of 5.8 m and a diameter of 2.5 m [1]. The closest detector to the interaction point, the Silicon Pixel detector, consists of the barrel pixel sub-detector (BPIX) composed of 4 layers and the forward pixel sub-detector (FPIX) with two forward endcaps, each composed of 3 disks, for a total of 1856 modules in the current configuration (Phase 1). Because of its proximity to the interaction point, the pixel detector is exposed to the highest density of tracks from the proton-proton collisions and, therefore, suffers more extensively the effects of the radiation damage. To tackle these effects, the pixel tracker was extracted from the CMS experimental cavern, underwent extensive repairs, was provided with a new innermost layer, and was reinstalled during the LHC Long Shutdown 2 (LS2). The silicon strip detector consists of 15 148 modules and is composed of the Tracker Inner Barrel (TIB), Tracker Inner Disks (TID), Tracker Outer Barrel (TOB), and Tracker EndCaps (TEC).

## 2. Track-based alignment

The tracker was specifically designed to allow for a very accurate determination of the trajectory of charged particles by ensuring an intrinsic spatial resolution of up to 10-30 microns. From the positions at a number of points registered in the detector in the form of hits, the trajectory of a charged particle can be reconstructed. Once the hits belonging to each trajectory are associated, a fit is performed and the track parameters such as the track curvature radius, impact parameter $d_{xy}$ in the xy plane, impact parameter $d_z$ along the beam direction, as well as the angles $\theta$ and $\phi$, are obtained. The tracking efficiency, a measure of how efficiently a charged particle passing through the tracker is reconstructed, provides a quantitative estimate of the tracking performance. Given the deflection of charged particles in a magnetic field, their transverse momentum can be computed as the product of the electric charge, the magnetic field, and the curvature radius. Thus, the track parameter uncertainties are propagated to the momentum measurement. Furthermore, the resolution of a reconstructed primary vertex position depends strongly on the number of tracks used to fit the vertex and the transverse momentum of those tracks. Hence, high-quality track reconstruction paves the way for accurate primary and secondary vertex reconstruction.

To ensure good tracking efficiency, as well as a precise momentum measurement and primary vertex reconstruction, the uncertainty on the track parameters needs to be reduced as much as possible. One of the main contributions to this uncertainty comes from the limited precision of the hit position measurements entering the track fit from which the track parameters are determined, which in turn is related to the overall limited knowledge of the position of the detector modules. The latter has two main contributions: the intrinsic spatial resolution of the sensors and the uncertainty related to the limited knowledge of the displacements, rotations, and surface deformations of the tracker modules. For an accurate determination of the track parameters, this second source of uncertainty in the position of the detector modules needs to be reduced to at least the intrinsic spatial resolution of the sensors. The correction of the position, orientation, and curvature of the tracker modules to reach a precision better than the intrinsic spatial resolution is the task of tracker alignment.

The so-called track-based alignment consists of fitting a set of tracks with an appropriate track model, and computing track-hit residuals, i.e., the difference between the measured hit position and the corresponding prediction obtained from the track fit. Geometry corrections can be derived from the $\chi^2$ minimization of these track-hit residuals. The Millepede and HipPy alignment algorithms are used by CMS to solve the $\chi^2$ minimization problem [2, 3]. The alignment parameters are determined with Millepede in a simultaneous fit of all tracks, involving two types of parameters: the local parameters that characterize the tracks used for the alignment, and nine global parameters that describe the position, orientation, and surface deformations of the modules. The local parameters of a single track are only connected to the subset of global parameters that are related to that track, and they are not directly connected to the local parameters of other tracks. The global parameters of each of the single modules of the detector can be corrected in a single alignment fit if enough tracks are available. On the other hand, when using the HipPy algorithm, the $\chi^2$ of each sensor is minimized with respect to a change in the local alignment of that sensor only, keeping the parameters of every other sensor fixed, in an iterative procedure.

Once the set of alignment constants is obtained, the improvement of post-alignment track-hits residuals is reviewed. Furthermore, before the new detector geometry is updated online for the data taking or used for the data reprocessing, the impact of the new set of alignment constants in the tracking performance, vertexing performance, and physics observables such as the mass of the Z boson resonance as function of the pseudorapidity and azimuthal angle is checked.

A simplified version of the offline alignment described above also runs online as part of the Prompt Calibration Loop (PCL). The PCL alignment uses the MillePede algorithm and performs the alignment of the pixel high-level structures at the level of ladders and panels, which ensures the consideration of radiation effects of the innermost layer of the barrel pixel detector already during data taking. The obtained constants are then used for the reconstruction of the next run if movements are above certain thresholds. Thus, the online and offline alignment are complementary components of the tracker alignment within CMS, one providing automated online correction of the pixel high-level structures and the other refining the alignment calibration with the possibility to reach each of the single modules of the detector in a single alignment fit.

## 3. Tracker alignment effort prior to the beginning of Run 3

The first data-taking exercise upon the restart of operations in 2021 consisted of recording cosmic ray muons with the magnetic field off, “cosmic run at zero Tesla” (CRUZET). The alignment with cosmic ray data has the advantage of allowing the update of the tracker alignment constants before the start of collision data taking. Major shifts in the pixel and strip sub-detectors (e.g., due to magnet cycles and temperature changes) can be identified and the geometry corrected accordingly before beams are injected into the collider and collision data becomes available. The very first alignment of the pixel detector after reinstallation in the experimental cavern was performed using 2.9M cosmic ray tracks recorded during the CRUZET period, at the level of single modules for the pixel detector and the outer barrel of the strip detector, and of half-barrels and half-cylinders for the rest of the strip partitions. This period was followed by cosmic data-taking with magnetic field at nominal value (3.8T), “cosmic run at four Tesla” (CRAFT). In this case, the geometry was derived using 765k cosmic ray tracks with the alignment corrections derived at the level of single modules for the barrel pixel and at the level of half-barrels and half-cylinders for the forward pixel and all of the strip sub-detectors. While the geometries derived with CRUZET and CRAFT data constituted relevant updates of the alignment constants starting from a potentially large misalignment, the results are statistically limited by the available number of cosmic ray tracks, especially in the forward pixel endcaps, and systematically limited by the lack of kinematic variety of the tracks sample. For a further improvement of the alignment calibration, a sample of 255.2M pp collision tracks, accumulated at a center-of-mass energy of 900 GeV and 3.8T magnetic field, was used. Finally, shortly before the start of pp collisions in 2022, the alignment was updated using 6.3M cosmic ray tracks recorded at 3.8 T. Alignment corrections were derived at the level of single modules for the pixel detector and at the level of half-barrels and half-cylinders for the different strip sub-detectors. A comparison of the performance of the different sets of alignment constants obtained with cosmic rays at 0T, cosmic rays at 3.8T, and pp collision tracks during 2021, as well as the alignment performance in 2022 prior to pp collisions at $\sqrt{s}$=13.6 TeV, are presented in this section.

### 3.1 Offline alignment using cosmic-ray and collision tracks (2021)

The distribution of the median of the track-hit residuals per module (DMRs) constitutes a measure of the tracking performance. The DMRs are monitored for all the tracker substructures, as shown for the barrel and forward pixel sub-detectors in Figure 1. A significant improvement on the track-hit residuals for the alignment with collision data over the alignments with cosmic ray muons only is observed. In the barrel region, DMR distributions can be obtained separately for the pixel barrel modules pointing radially inwards or outwards. The difference of their mean values $\Delta\mu$ in the local-x (x') direction shown in Figure 2 as a function of the delivered integrated luminosity constitutes a measure of the reduction of Lorentz drift angle effects with the alignment procedure.

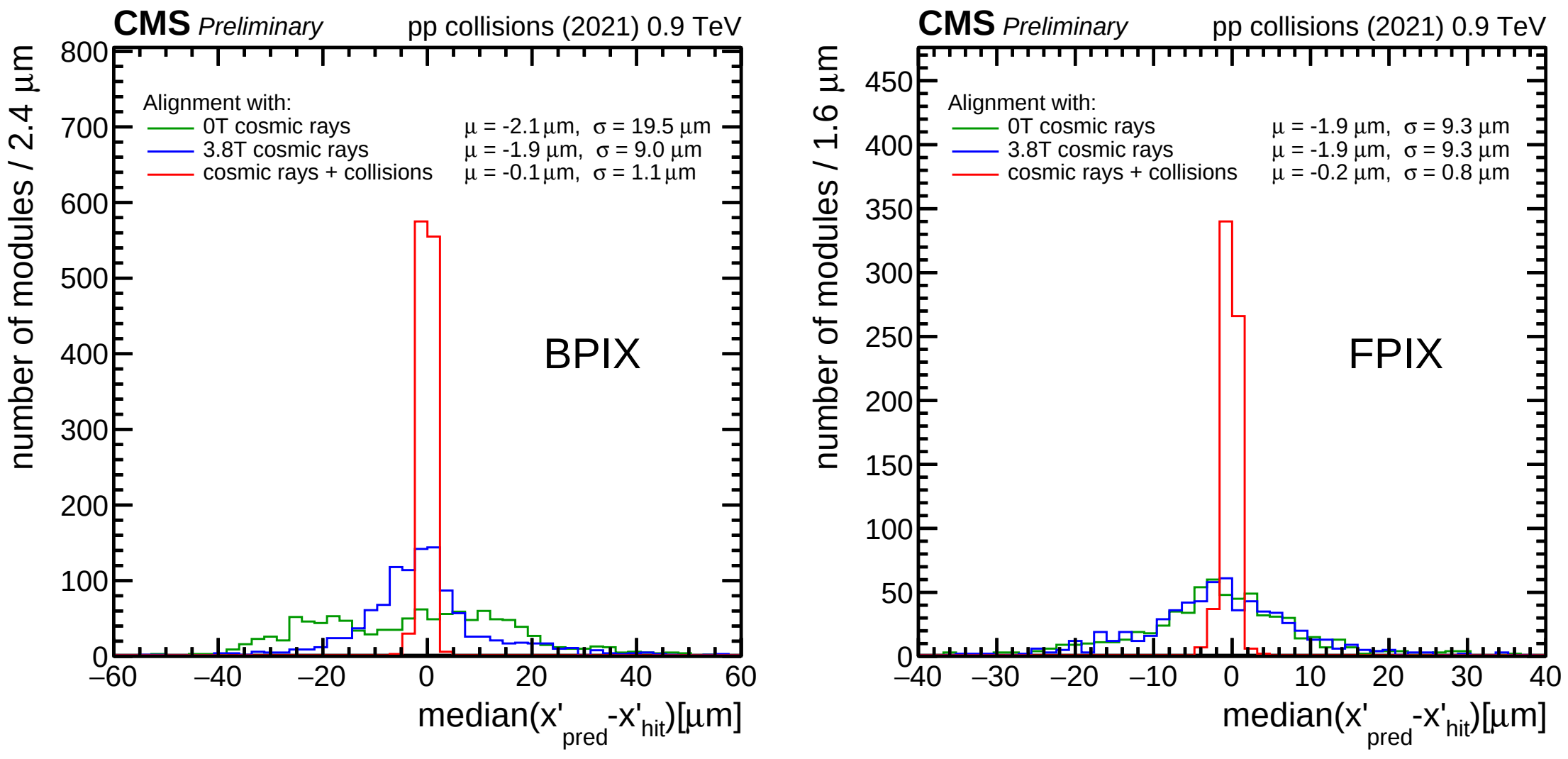

**Figure 1:** The distribution of median residuals is shown for the local-x coordinate in the barrel pixel (left) and forward pixel (right). The alignment constants used for the reprocessing of the pp collision data (red) are compared with the ones derived using cosmic rays only, recorded at 0T (green) and 3.8T (blue). The quoted means $\mu$ and standard deviations $\sigma$ correspond to parameters of a Gaussian fit to the distributions [4].

The effect of the alignment calibration on the reconstruction of physics objects is also studied. The distance between tracks and the unbiased track-vertex residuals is studied, searching for potential biases in the primary vertex reconstruction. The mean value of the unbiased track-vertex residuals is shown in Figure 3 for the longitudinal and transverse planes, with a significant reduction of the bias when collision tracks are included in the alignment procedure.

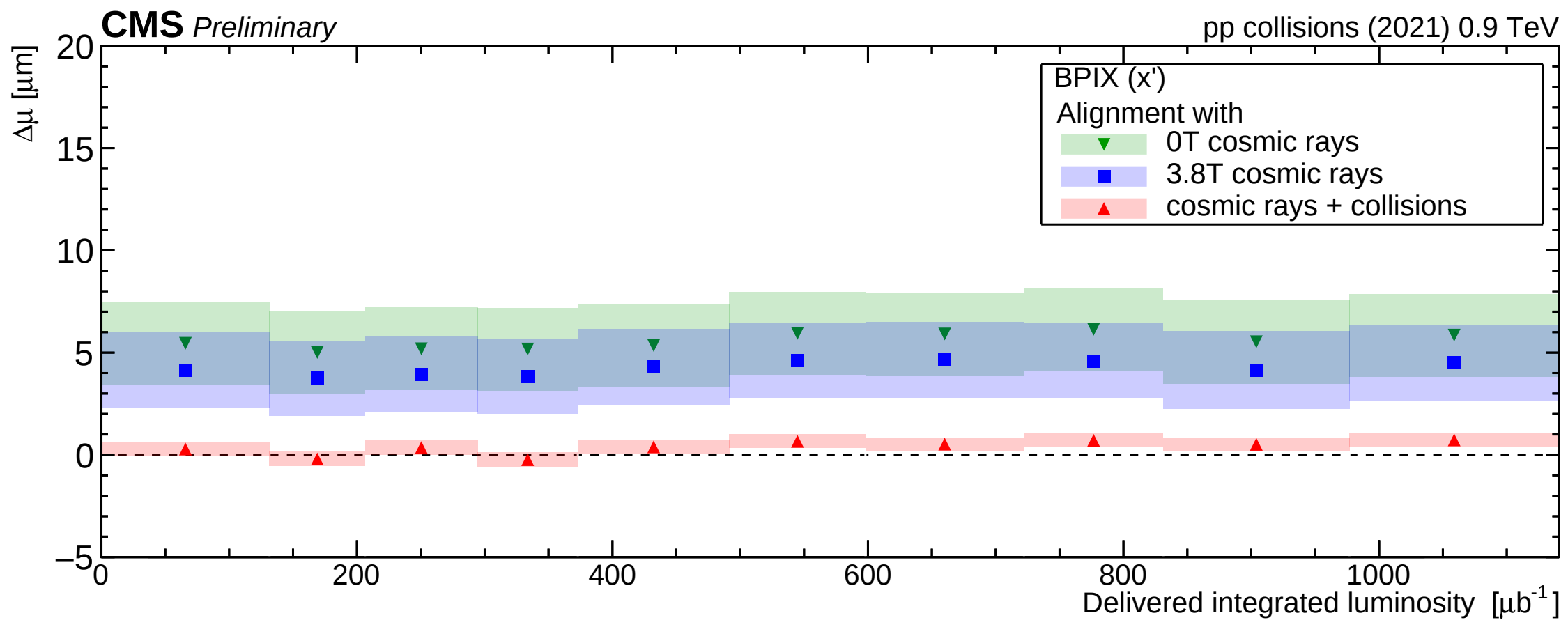

**Figure 2:** Difference between the mean values $\Delta\mu$ obtained separately for the modules with the electric field pointing radially inwards or outwards. After alignment with cosmic and collision tracks, the mean difference $\Delta\mu$ is consistently closer to zero [4].

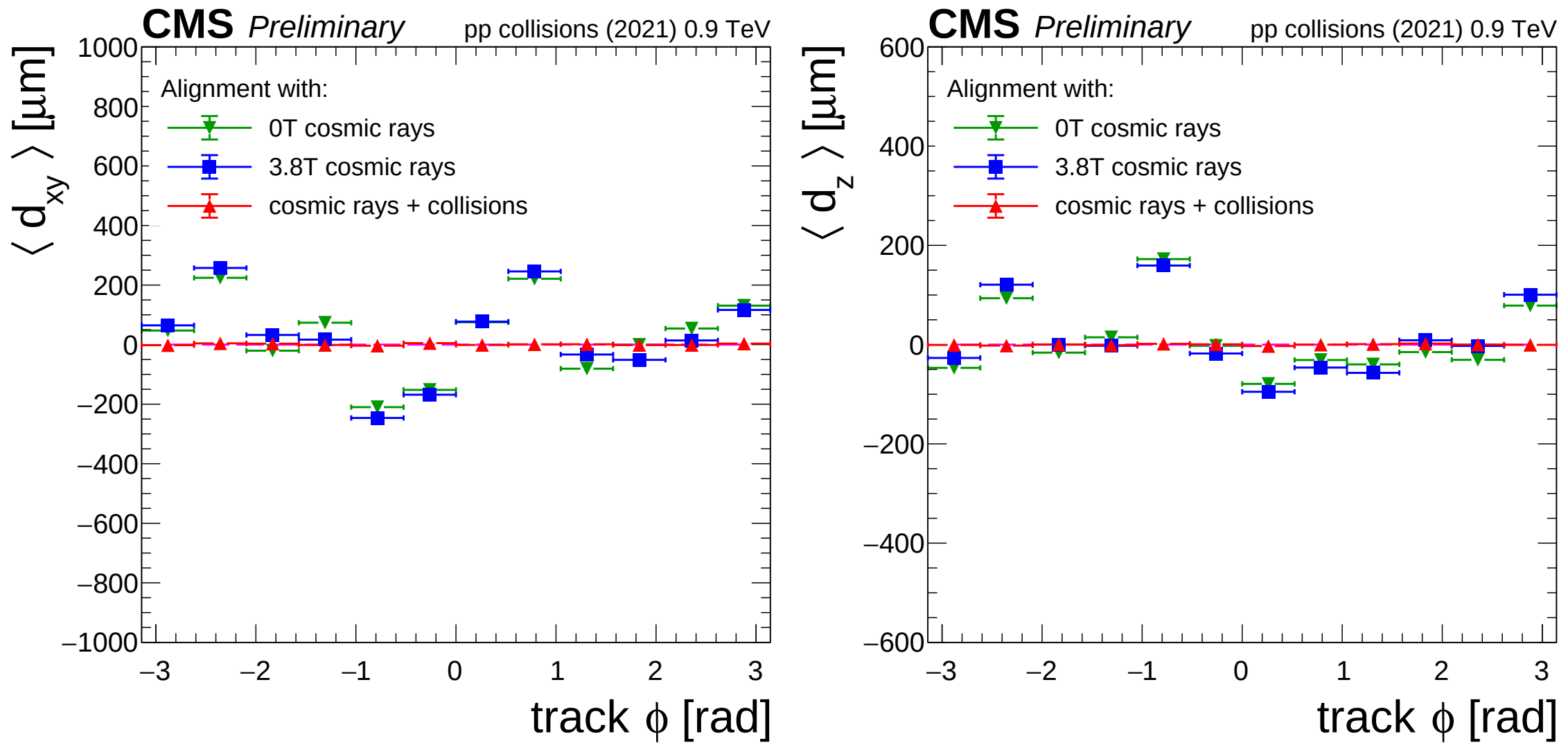

**Figure 3:** The mean track-vertex impact parameter in the transverse $d_{xy}$ plane (left) and longitudinal $d_z$ plane (right) in bins of the track azimuthal angle $\phi$ is shown [4].

## 4. Offline alignment using cosmic-ray tracks (2022)

The alignment constants were updated before the start of Run 3 using cosmic ray muons recorded at 3.8 T, to correct for movements caused by the magnet cycle during the 2021-2022 winter break and repeated temperature cycles due to strip detector maintenance. After the geometry update, the bias on the distribution of median residuals for the forward pixel detector was corrected, as shown in Figure 4, left. Furthermore, the difference in the track impact parameters in the transverse plane $d_{xy}$ for cosmic ray tracks passing through the pixel detector and split into two halves at their point of closest approach to the interaction region was also reduced, as shown in Figure 4, right.

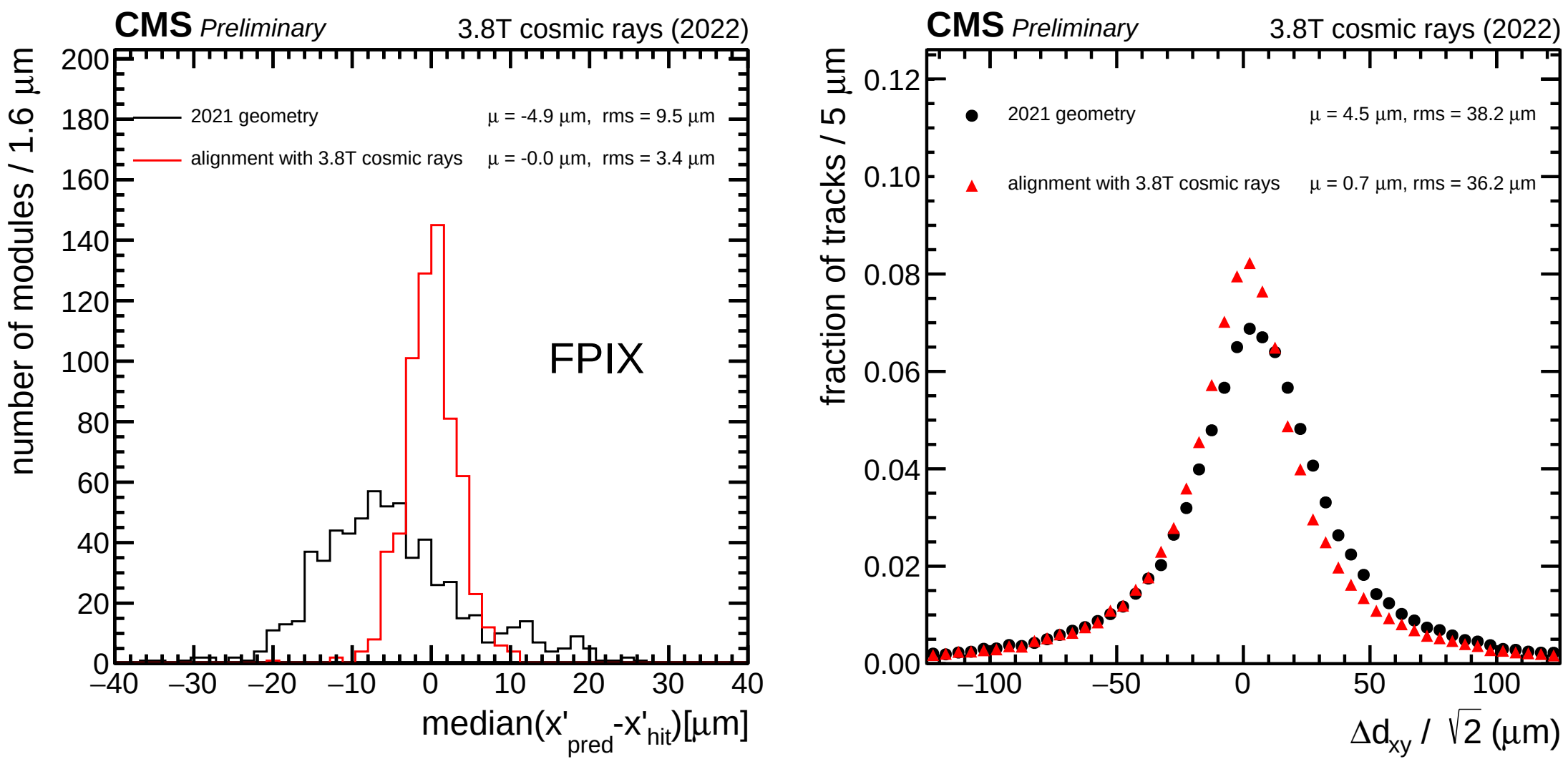

**Figure 4:** Distribution of median residuals for the local-x coordinate in the forward pixel (left) and difference in track impact parameters in the transverse plane $d_{xy}$ (right) [5].

## 5. Summary

The tracker alignment effort during the Run 3 commissioning period has been presented. The online alignment in the Prompt Calibration Loop and the strategy followed for the alignment calibration considering the availability of tracks with certain topologies have been discussed. Finally, the data-driven methods used to derive the alignment parameters and the set of validations that monitor the physics performance after the update of the tracker alignment constants have been presented.

## References

[1] CMS Collaboration, *The CMS Experiment at the CERN LHC*, 2008 JINST 3 S08004, doi:10.1088/1748-0221/3/08/S08004.

[2] V. Blobel and C. Kleinwort, *A new method for the high-precision alignment of track detectors*, Proceedings of Conference on Advanced Statistical Techniques in Particle Physics, Durham, UK, 2002, https://inspirehep.net/literature/589639.

[3] CMS Collaboration, *Strategies and performance of the CMS silicon tracker alignment during LHC Run 2*, 2022 Nucl. Instrum. Methods A 1037 166795, doi:10.1016/j.nima.2022.166795.

[4] CMS Collaboration, *CMS Tracker Alignment Performance Results CRAFT 2022*, CMS Status Report, 150th LHCC Meeting - OPEN Session, 2022, https://indico.cern.ch/event/1156732/.

[5] CMS Collaboration, *Tracker Alignment Performance in 2021*, CMS-DP-2022/017, 2022, https://cds.cern.ch/record/2813999/.